\documentclass{article}

 \usepackage[preprint]{neurips_2026}

\usepackage[utf8]{inputenc} 
\usepackage[T1]{fontenc}    
\usepackage{hyperref}       
\usepackage{url}            
\usepackage{booktabs}       
\usepackage{amsfonts}       
\usepackage{nicefrac}       
\usepackage{microtype}      
\usepackage{xcolor}         
\usepackage{multirow}       
\usepackage{graphicx}
\usepackage{float}
\usepackage{fancyvrb}

\title{SWE Context Bench: A Benchmark for Context Learning in Coding}

\author{%
  {\bf
  Jiayuan Zhu\textsuperscript{1},
  Junde Wu\textsuperscript{2},
  Minhao Hu\textsuperscript{2},
  Shengda Zhu\textsuperscript{3},
  Jiazhen Pan\textsuperscript{4},
  Weixiang Shen\textsuperscript{4}} \\
  {\bf
  Yijun Yang\textsuperscript{3},
  Fenglin Liu\textsuperscript{2},
  Jianye Hao\textsuperscript{5},
  Yueming Jin\textsuperscript{6},
  Qirong Ho\textsuperscript{7},
  Min Xu\textsuperscript{7,8}} \\
  \textsuperscript{1}Independent Researcher \quad \textsuperscript{2}University of Oxford \quad \textsuperscript{3}University of Edinburgh \\
  \textsuperscript{4}Technical University of Munich \quad \textsuperscript{5}MemoraX AI \quad \textsuperscript{6}National University of Singapore \\
  \textsuperscript{7}Mohamed bin Zayed University of Artificial Intelligence \quad \textsuperscript{8}Carnegie Mellon University
}

\begin{document}

\maketitle

\begin{abstract}
Large language models are increasingly used as coding agents for software engineering tasks. Current benchmarks mainly evaluate whether the agent can correctly solve the request or fix the bugs. They largely treat tasks as independent and do not assess whether agents can reuse previous experience across related problems. As a result, the efficiency gains from reusing the previous experience remains difficult to measure. We introduce SWE-ContextBench, a benchmark designed to explicitly evaluate context understanding and retrieval in coding agents. SWE-ContextBench consists of 1,100 base tasks with another 376 related tasks derived from real dependency and reference relationships among GitHub issues and pull requests. SWE-ContextBench groups base tasks and related tasks with shared context across 51 unique repositories and 9 programming languages. The benchmark evaluates how accurately and efficiently agents solve related issues when prior cases are available in context. Using SWE-ContextBench, we study the behavior of multiple coding agents across varying context reuse settings and retrieval strategies. Our results show that accurately summarized and retrieved previous experience can significantly improve resolution accuracy and reduce runtime and token cost, particularly on harder tasks.  In contrast, unfiltered or incorrectly selected context provides limited or negative benefits. These findings highlight the importance of context management and retrieval accuracy, and position SWE-ContextBench as a principled benchmark for studying context learning in coding agents.

\end{abstract}

\section{Introduction}
Large language models (LLMs) are increasingly deployed as coding agents, supporting tasks such as code generation, bug fixing, and navigation of complex code repositories. Early code benchmarks primarily evaluated isolated synthesis tasks and focused on functional correctness. However, these benchmarks have become saturated, with near perfect performance on HumanEval \cite{chen2021evaluating} and MBPP \cite{austin2021program}. To address this limitation, recent benchmarks extend evaluation to repository level software engineering settings, where models reason over larger codebases and generate code changes evaluated in realistic repository contexts \cite{liu2306repobench}.

\begin{figure*}[hbt!]
\vspace{-30pt}
    \centering
    \includegraphics[width=0.9\linewidth]{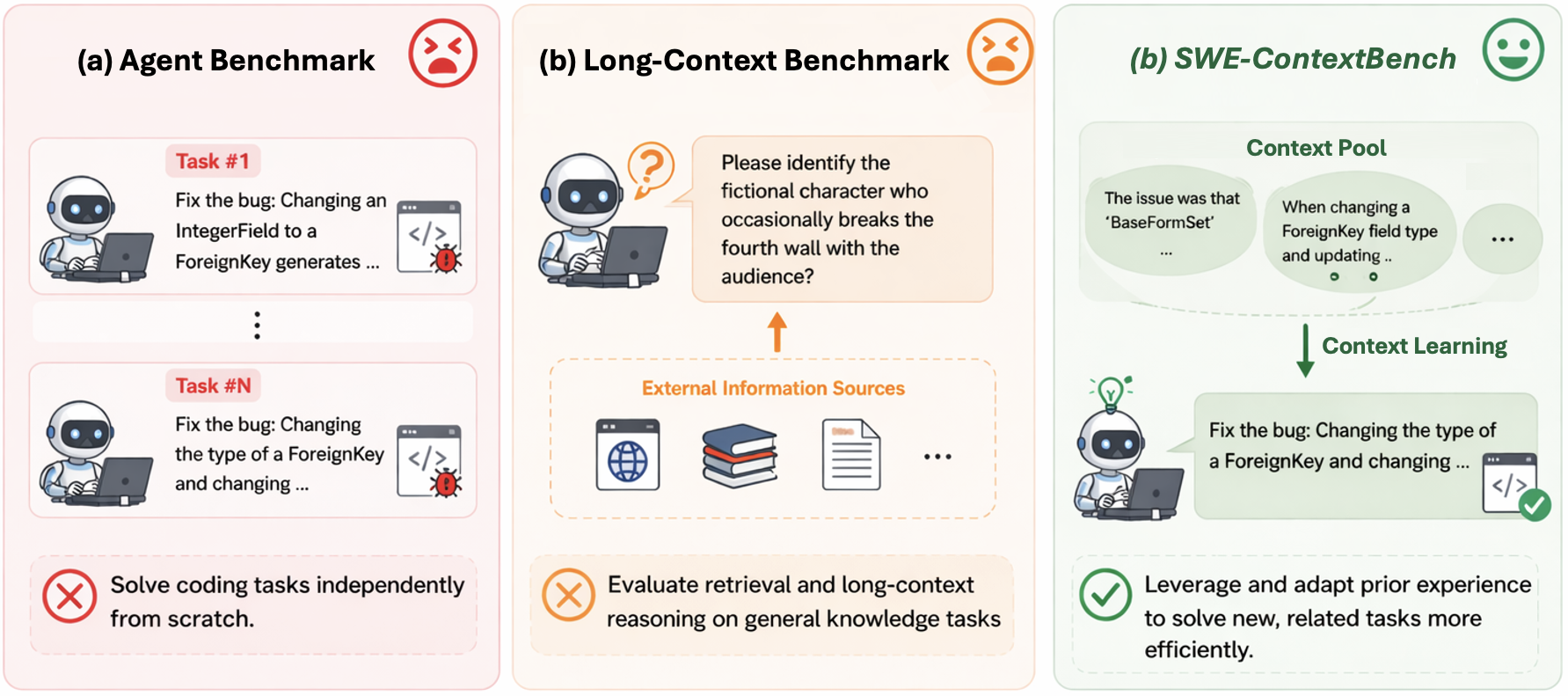} 
    \vspace{-5pt}
    \caption{Compare SWE-ContextBench with existing agent benchmark and long-context benchmark. }
    \label{fig:overview}
\vspace{-15pt}
\end{figure*}

Despite these advances, existing agent benchmarks remain focused on solving individual tasks from scratch (Figure \ref{fig:overview} (a)). They primarily evaluate knowledge stored in model weights \cite{yang2024swe, zeng2025glm} and short term context within the input window \cite{yu2025memagent, zhou2025mem1}, but do not assess how agents learn from and reuse past experience across tasks. As a result, each task is treated as an independent episode, preventing explicit evaluation of context learning and its impact on efficiency. Prior studies suggest that incorporating memory of past interactions enables agents to learn from previous mistakes and generalize knowledge across related tasks \cite{chhikara2023knowledge}. This observation aligns with real world software development, where recurring problem patterns and incremental refinement are common, and engineers routinely learn from past successes and failures.

In parallel, many recent agent models increasingly incorporate external context augmentation by retrieving relevant information beyond the current task trajectory \cite{chhikara2025mem0, li2025memos, xu2025mem}. However, existing long-context and memory-oriented benchmarks still fall short of addressing context learning in programming tasks. Benchmarks such as BrowseComp \cite{wei2025browsecomp}, WideSearch \cite{wong2025widesearch}, GAIA \cite{mialon2023gaia}, and LoCoMo \cite{maharana2024evaluating} are designed to assess the ability to retrieve and reason over large amounts of information within a single task (Figure \ref{fig:overview} (b)). As a result, they measure retrieval to relevant memory rather than the ability to summarize and reuse experience across tasks. Consequently, they fail to capture whether agents can leverage prior problem-solving history to improve performance on related future tasks.

Taken together, these limitations make the context learning capabilities of coding agents difficult to quantify from existing agent centric and context based benchmarks. This gap highlights the need for a benchmark that explicitly evaluates context learning across tasks. An effective coding agent should not only produce correct solutions, but also exploit prior experience to solve related tasks more efficiently. To this end, we propose evaluating agents along multiple dimensions: task accuracy variation, context aware solving time, and token efficiency. The latter two measure the ability to locate and reuse relevant prior tasks and to reduce overall computational cost through previous reasoning. 

In this paper, we propose \textbf{SWE-ContxetBench}, a benchmark designed to study context learning in software engineering agents, as illustrated in Figure \ref{fig:overview} (c). SWE-ContextBench is built upon SWE-Bench Lite, SWE-Bench Multilingual, and SWE-Bench Verified \cite{jimenez2024swebench}, where related instances are carefully selected by humans. In total, SWE-ContextBench contains 1,476 tasks from 51 unique GitHub repositories spanning 9 programming languages, including 1,100 base tasks and 376 related tasks. To support faster and more cost effective evaluation, we also introduce SWE-ContextBench Lite, which consists of 300 base tasks and 99 related tasks. The benchmark enables direct evaluation of how effectively agents can learn from previous context when solving new tasks. As such, SWE-ContextBench provides a principled and realistic testbed for studying context learning in coding agents, complementing existing programming and context benchmarks. Our study shows that accurate summarization and retrieval of past experience significantly improve resolution accuracy while reducing runtime and token costs. However, models do not always leverage this information effectively and can be misled by incorrect prior assumptions. Retrieval alone is therefore insufficient; agents must also learn when to trust and how to adapt retrieved context for new tasks.

\section{SWE-ContxetBench}

\subsection{Desiderata: What metrics should an ideal benchmark include?}
An ideal benchmark for coding agents should evaluate not only correctness, but also how effectively prior context is reused to improve efficiency. We identify three key dimensions. 1) \textbf{Accuracy Variation.} Accuracy variation measures how an agent's performance changes when prior context is available. Instead of evaluating correctness in isolation, it quantifies the improvement or degradation in solution accuracy relative to a no-context baseline. We ground this evaluation in well-maintained GitHub repositories, where issues have human-verified fixes and are validated by test suites. Newly added or modified tests help determine whether a solution successfully resolves the issue. This setup enables to directly assess whether access to previous context leads to measurable gains in task performance. 2) \textbf{Time efficiency.} This measures whether an agent solves tasks faster by leveraging prior context. Given previously solved related issues, an effective agent should identify relevant cases and learn from them to reduce solving time, which can be measured through wall clock time or search behavior. 3) \textbf{Cost efficiency.} This captures the computational resources required to solve a task. By successfully learning from prior context, agents can skip redundant reasoning, reducing token usage and overall cost. Token consumption thus serves as a practical and quantitative proxy for efficiency gains. Together, these three dimensions provide a principled framework for evaluating coding agents.

\subsection{Data Construction}
\subsubsection{Base Dataset Selection}
\begin{figure*}[hbt!]
\vspace{-5pt}
    \centering
    \includegraphics[width=0.85\linewidth]{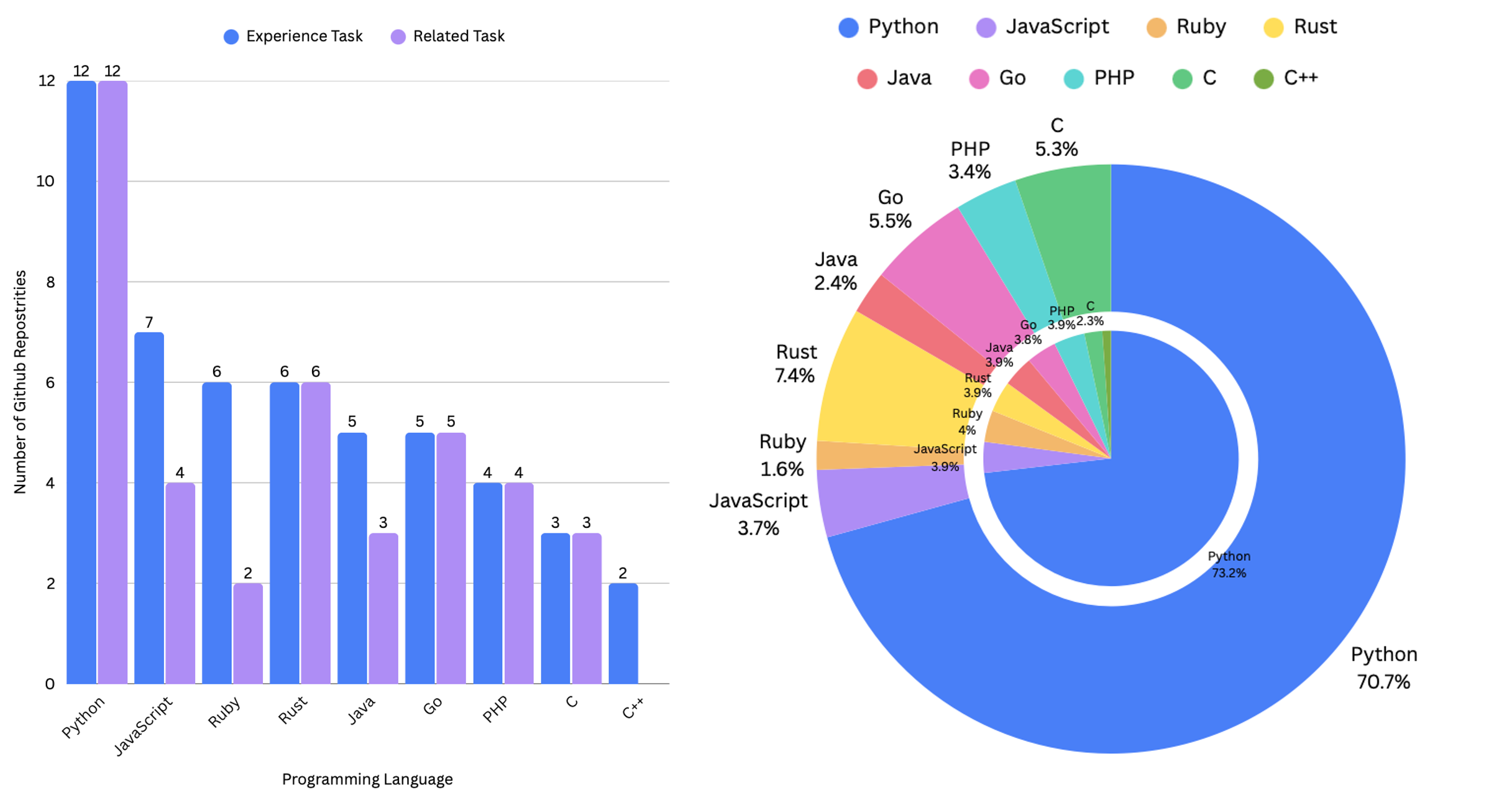} 
    \vspace{-5pt}
    \caption{Distribution of programming languages across tasks. The bar chart shows the number of repositories per language for experience and related tasks. The donut chart shows the percentage distribution by language, with the inner ring for experience tasks and the outer ring for related tasks.}
    \label{fig:repo_analysis}
\vspace{-10pt}
\end{figure*}

We build our benchmark on SWE-Bench Lite, SWE-Bench Multilingual, and SWE-Bench Verified \cite{jimenez2023swe}. It originally contains 1,100 instances from 51 real-world GitHub repositories across 9 programming language, including Python, JavaScript, Ruby, Rust, Java, Go, PHP, C, and C++ (Figure \ref{fig:repo_analysis}). Each instance includes a problem description from GitHub issue and the corresponding code changes from pull requests, validated by test suites. This setup preserves realistic repository complexity and task difficulty, making it well suited for controlled agent evaluation at scale. 

\subsubsection{Context Trajectory Collection} \label{sec:experience_traj}
For each instance in the base dataset, we run a base coding agent and record its full trajectory, including tool calls, file navigation, and intermediate reasoning steps. These trajectories are stored as past contexts and form a context pool that agents can later retrieve when solving related tasks. This enables explicit evaluation of whether agents learn from prior context instead of solving from scratch.

\subsubsection{Identification of Related Task Instances} \label{sec:task_instance_type}
\begin{figure*}[hbt!]
\vspace{-5pt}
    \centering
    \includegraphics[width=0.85\linewidth]{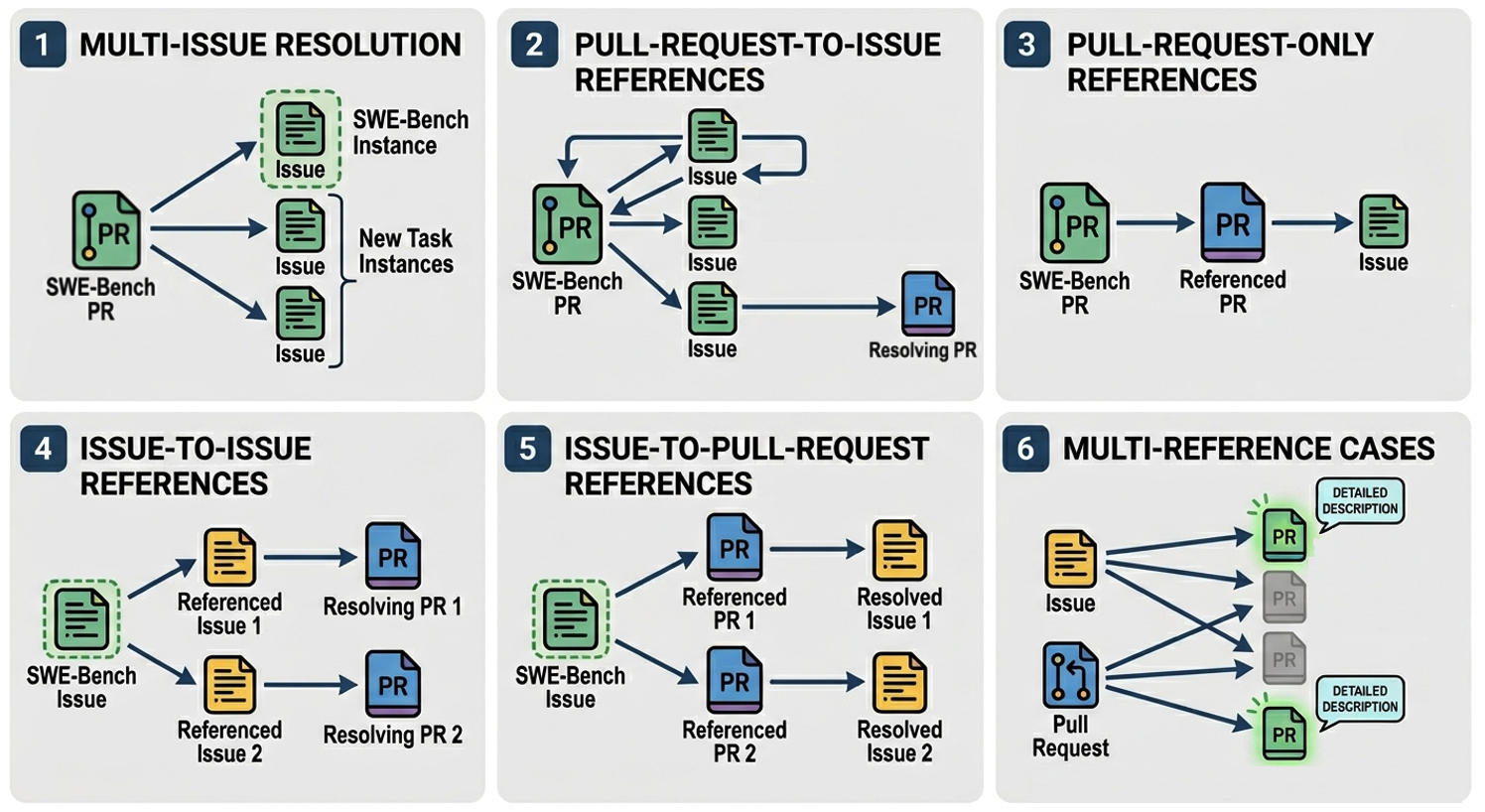} 
    \vspace{-5pt}
    \caption{Six workflows for expanding SWE-Bench base dataset by identifying related task instances through issue and pull request cross-references.}
    \label{fig:related_task_type}
\vspace{-10pt}
\end{figure*}

To construct task sequences with shared context, we manually analyze the pull requests and issues for each base dataset instance. For each instance, we check whether the corresponding pull request or issue references other issues or pull requests and verify their resolution relationships. For each related case, we extract the issue description and the associated pull request that introduces the code changes. Specifically, we consider the following as related task instances to the base dataset (Figure \ref{fig:related_task_type}):

\begin{itemize}
    \item \textbf{Multi-issue resolution}: A single pull request resolves multiple issues, but only one appears in base dataset. Others are treated as additional task instances paired with same pull request.

    \item \textbf{Pull-request-to-issue references}: The pull request of a base dataset instance references other issues. If resolved by the same pull request, they form related task instances. If resolved by different pull requests, each issue is paired with its corresponding resolver.
    
    \item \textbf{Pull-request-only references}: A base dataset instance's pull request references another pull request. The issue resolved by the referenced pull request is used as a valid task instance.
    
    \item \textbf{Issue-to-issue references}: An issue references other issues. Each referenced issue is paired with its resolving pull request to form a related task instance.
    
    \item \textbf{Issue-to-pull-request references}: An issue references one or more pull requests. The resolved issues associated with those pull requests are used to construct related task instances.
    
    \item \textbf{Multi-references}: An issue or pull request references other pull requests. Those with sufficient descriptions are treated as task instances, using their descriptions as issue descriptions.
\end{itemize}

Across the 1,100 base instances, we identify 319 such interdependent task instances. These subtypes reflect common software development patterns. Each instance is manually verified to ensure that it is a meaningful software engineering task with clear relationships and identifiable resolution.

\subsubsection{Recursive Context Expansion}
Starting from the 319 identified instances, we apply the same rules to each new issue or pull request. By examining their discussions, we identify additional references. This recursive process yields 57 additional interdependent task instances, forming new task sequences for evaluating context learning.

\subsubsection{Resulting Benchmark Tasks}
The final benchmark consists of related software engineering tasks derived from real-world repositories. It includes 1,100 \emph{experience tasks}, whose solution trajectories form a reusable context pool, and 376 \emph{related tasks} identified through explicit dependency and reference analysis. In total, SWE-ContextBench contains 1,476 tasks across 51 repositories, with details in Table \ref{tab:source_stat}.

\begin{table}[H]
\vspace{-5pt}
\centering
\caption{Breakdown of SWE-ContextBench by source benchmark, showing counts of experience/related repositories and experience/related tasks, plus total tasks.}
\vspace{-4pt}
\resizebox{0.8\linewidth}{!}{
\begin{tabular}{c|ccccc}
\hline \hline
                       & \begin{tabular}[c]{@{}c@{}}Experience\\ Repos\end{tabular} & \begin{tabular}[c]{@{}c@{}}Related\\ Repos\end{tabular} & \begin{tabular}[c]{@{}c@{}}Experience\\ Tasks\end{tabular} & \begin{tabular}[c]{@{}c@{}}Related\\ Tasks\end{tabular} & \begin{tabular}[c]{@{}c@{}}Overall\\ Tasks\end{tabular} \\ \hline
SWE-Bench Lite         & 12                                                   & 12                                                      & 300                                                  & 99                                                      & 399                                                     \\
SWE-Bench Multilingual & 39                                                   & 27                                                      & 300                                                  & 111                                                     & 411                                                     \\
SWE-Bench Verified     & 12                                                   & 12                                                      & 500                                                  & 166                                                     & 666                                                     \\ \hline
SWE-ContextBench       & 51                                                   & 39                                                      & 1,100                                                & 376                                                     & 1,476      \\ \hline \hline                                            
\end{tabular}
}\label{tab:source_stat}
\vspace{-13pt}
\end{table}

Additionally, Figure \ref{fig:repo_analysis} shows programming-language-level statistics across both task sets. SWE-ContextBench covers 9 languages in experience tasks and 8 in related tasks, with Python dominating in both (12 vs. 12 repositories; 73.2\% vs. 70.7\% of tasks). Other languages show similar distributions, including JavaScript, Rust, and Go, with only minor differences across categories. Overall, both the repository counts and percentage distributions remain closely aligned, indicating that the related-task set preserves the language composition of the experience-task pool.

The 376 related tasks are derived from common software development patterns described in Section~\ref{sec:task_instance_type}. Specifically, we identify 46 tasks from \textbf{multi-issue resolution}. From \textbf{pull-request-to-issue references}, we find 72 tasks, including 42 issues resolved by the same pull request and 30 by different ones. We identify 34 tasks from \textbf{pull-request-only references}. From \textbf{issue-to-issue references}, we obtain 103 tasks (78 issues resolved by their own corresponding pull requests and 25 issues sharing the same pull request as the original issue). We also identify 41 tasks from \textbf{issue-to-pull-request references}, of which 16 correspond to a single new resolved issue and 25 referencing multiple pull requests. Finally, we identify 23 tasks from \textbf{multi-references}. Beyond these first-order relations, \textbf{recursive context expansion} yields 57 additional related tasks, capturing second-order dependencies.

By preserving real codebases, issue descriptions, and validated code changes, the benchmark enables direct evaluation of whether agents can learn from previous context for new but related coding tasks.

\subsubsection{Resulting Benchmark Structure}
Context pool contains solution trajectories from 1,100 \emph{experience tasks} (Section~\ref{sec:experience_traj}). For 376 \emph{related tasks}, the dataset includes two components: task construction and test-based evaluation data.

\textbf{Task construction.}
Each task defines a software bug-fixing problem, specifying where to start, what issue to resolve, and how correctness is evaluated. Each related task is derived from a GitHub pull request and includes the repository state before the fix (\texttt{base\_commit}), the problem statement, and the resolving code changes. The \texttt{base\_commit} is the base branch SHA, representing pre-fix state. 

Tasks are constructed through an automated pipeline via the GitHub API. Problem statements are extracted from GitHub issue titles and descriptions, or Django Trac ticket system for Django repositories. The ground-truth solution is taken from the pull request’s unified diff and split into a \texttt{test\_patch} and a \texttt{solution\_patch}. The \texttt{test\_patch} contains changes to test files, identified by keywords like “test”, “tests”, \textit{etc}. The \texttt{solution\_patch} contains changes to implementation files.

\textbf{Test-based evaluation.}
For evaluation, we construct a rigorous test-based validation framework. Starting from the \texttt{base\_commit}, we set up an isolated environment with the detected dependencies and run the existing test suite to obtain baseline results (\texttt{results\_before}). We then apply the \texttt{test\_patch}, which introduces tests that specify the expected behavior, followed by the \texttt{solution\_patch}. After applying both patches, we re-run the test suite to obtain \texttt{results\_after}. 

Evaluation is based on two test sets. The \texttt{FAIL\_TO\_PASS} set includes tests that fail, error, or are absent in \texttt{results\_before} but pass in \texttt{results\_after}, indicating the issue is fixed. The \texttt{PASS\_TO\_PASS} set includes tests that pass both before and after, ensuring no regressions. Together, they ensure that a valid solution both fixes the intended problem and preserves existing correct behavior.

\begin{table}[H]
\vspace{-5pt}
\centering
\caption{Language-level average numbers of \texttt{FAIL\_TO\_PASS} and \texttt{PASS\_TO\_PASS} tests across the 376 related tasks, along with their overall averages.}
\vspace{-5pt}
\resizebox{0.92\linewidth}{!}{
\begin{tabular}{c|cccccccc|c}
\hline \hline
               & Python & JavaScript & Ruby  & Rust  & Java   & Go    & PHP  & C      & Overall \\ \hline
FAIL\_TO\_PASS & 4.3    & 2.9        & 19.2  & 5.6   & 2.6    & 9.7   & 6.3  & 2.6    & 4.8     \\
PASS\_TO\_PASS & 145.0  & 303.7      & 172.7 & 534.3 & 1219.8 & 201.5 & 59.7 & 243.9  & 211.5   \\ \hline
Overall        & 149.3  & 306.6      & 191.9 & 539.9 & 1222.4 & 211.2 & 66.0 & .246.5 & 216.3  \\ \hline \hline
\end{tabular}
}\label{tab:language_level_tests}
\vspace{-14pt}
\end{table}

On average, each of the 376 related tasks contains 4.8 \texttt{FAIL\_TO\_PASS} tests and 211.5 \texttt{PASS\_TO\_PASS} tests. This indicates that tasks are typically validated by a small set of tests for the fix and a larger set for regression. Table~\ref{tab:language_level_tests} reports language-level averages for both test types.

\subsection{SWE-ContxetBench Lite}
To support faster and cost-effective evaluation of context learning, we release SWE-ContextBench Lite. Built from SWE-Bench Lite, it contains 300 base tasks for the context pool and 99 related tasks for evaluating (399 tasks in total). It focuses on Python tasks across 12 repositories.

\section{Experiment}
\subsection{Experimental Setup}
To build the context pool, we run Claude Code (Claude Sonnet~4.5) on the 1,100 \emph{experience tasks}. For each task, the agent is given only the repository name, base commit, and problem statement. The full interaction trajectory, including file navigation and intermediate reasoning, is recorded as reusable context. All \emph{experience tasks} are executed independently in isolated environments to prevent information leakage across tasks. In addition, the quality of the \emph{experience tasks} is not explicitly controlled or filtered to reflect real-world software engineering environments (Section~\ref{sec:appendex_experience_pool_quality}).

For the 376 \emph{related tasks}, each is also run independently in a fresh environment. Context learning is enabled only through explicit retrieval from the context pool, with no implicit state carryover between runs. For each task, the agent receives the repository name, base commit, and problem statement (Section~\ref{sec:appendix_instruction}). A clean testbed is created by cloning the repository, checking out the base commit, and installing dependencies. The agent generates code changes (excluding test files), which are saved as a unified diff. The testbed is then removed, ensuring isolated and reproducible evaluation.

\subsection{Performance across Five Coding Agents}
\subsubsection{Accuracy Analysis}
Table~\ref{tab:all_base_agents} compares five coding agents on accuracy under two settings: a baseline without retrieved context and an oracle summary setting using gold relevant summaries from the context pool. Each summary averages 217.1 tokens and is taken from the final summary paragraph of the corresponding trajectory. This oracle setting isolates the impact of high-quality context reuse without retrieval errors. 

Overall, closed-source agents achieve the highest accuracy, and oracle summaries improve performance for most models. In the baseline, GPT-5.3 Codex performs best on function-level localization (19.25\%), line-level localization (14.71\%), and task resolution (22.60\%), showing strong localization and code fix ability. Among open-weight models, MiniMax~2.7 achieves the highest resolution rate (19.68\%), while Qwen3.5-plus and GLM-5 remain competitive on localization but lag in end-to-end success. With oracle summaries, most agents improve: GPT-5.3 Codex achieves the best file localization (57.22\%) and the best overall resolution rate (23.94\%), Claude Sonnet~4.5 increases to 23.40\% resolved, and MiniMax~2.7 rises to 22.07\%, narrowing the gap with the strongest proprietary systems.

\begin{table}[H]
\centering
\vspace{-6pt}
\caption{Performance of five coding agents on SWE-ContextBench under baseline and oracle-summary settings. We report localization-based accuracy metrics together with task resolution rate, and efficiency metrics including tool calls, token usage, runtime, and API cost.}
\vspace{-4pt}
\resizebox{0.95\linewidth}{!}{
\begin{tabular}{cccccccccc}
                    \hline\hline
                  & \multicolumn{1}{l}{}
                  & \multicolumn{4}{c}{\textbf{Accuracy Analysis ($\uparrow$})}                                                                                                                                                                                                                                          & \multicolumn{4}{c}{\textbf{Efficiency Analysis ($\downarrow$)}}                                                                                                                                                                                     \\
                  & \begin{tabular}[c]{@{}c@{}}Agent \\ Layer\end{tabular}
                  & \begin{tabular}[c]{@{}c@{}}File Correct\\ Location (\%)\end{tabular} & \begin{tabular}[c]{@{}c@{}}Function Correct\\ Location (\%)\end{tabular} & \begin{tabular}[c]{@{}c@{}}Line Correct\\ Location (\%)\end{tabular} & \begin{tabular}[c]{@{}c@{}}Resolved \\ (\%)\end{tabular} & \begin{tabular}[c]{@{}c@{}}Avg Tool\\ Calls\end{tabular} & \begin{tabular}[c]{@{}c@{}}Avg\\ Tokens\end{tabular} & \begin{tabular}[c]{@{}c@{}}Avg\\ Time (s)\end{tabular} & \begin{tabular}[c]{@{}c@{}}Cost\\ (\$)\end{tabular} \\ \hline \hline
\multicolumn{9}{l}{\textit{Closed-source Baseline}}                                                                                                                                                                                                                                                                                                                                                                                                                                                                                    \\
Claude Sonnet 4.5 & Claude Code & 52.14                                                                & 14.97                                                                    & 11.76                                                                & 19.68                                                    & 44.67                                                    & 1,701,548.05                                         & 344.47                                                 & 0.67                                                \\
GPT-5.3 Codex   & Codex   & 54.01                                                                & \textbf{19.25}                                                           & \textbf{14.71}                                                       & \textbf{22.60}                                           & 28.91                                                    & 858,252.71                                           & 349.41                                                 & 0.31                                                \\ 
\multicolumn{9}{l}{\textit{Open-weight Baseline}}                                                                                                                                                                                                                                                                                                                                                                                                                                                                                      \\
Qwen3.5-plus   & Qwen Code    & 52.96                                                                & 16.94                                                                    & 12.10                                                                & 16.22                                                    & 52.10                                                    & 2,332,845.20                                         & 523.47                                                 & 0.33                                                \\
GLM-5    & Qwen Code          & 53.89                                                                & 16.89                                                                    & 12.60                                                                & 18.35                                                    & 45.32                                                    & 1,933,888.48                                         & 800.99                                                 & 0.58                                                \\
MiniMax 2.7  & Claude Code     & 52.41                                                                & 17.11                                                                    & 12.83                                                                & 19.68                                                    & 53.24                                                    & 1,615,255.05                                         & 691.28                                                 & 0.65                                                \\ \hline \hline
\multicolumn{9}{l}{\textit{Closed-source Oracle Summary}}                                                                                                                                                                                                                                                                                                                                                                                                                                                                              \\
Claude Sonnet 4.5 & Claude Code & 53.21                                                                & 15.78                                                                    & 11.76                                                                & 23.40                                                    & 36.19                                                    & 1,352,829.76                                         & 280.62                                                 & 0.55                                                \\
GPT-5.3 Codex & Codex    & \textbf{57.22}                                                       & 18.18                                                                    & 13.64                                                                & \textbf{23.94}                                           & \textbf{27.41}                                           & \textbf{718,919.86}                                  & \textbf{362.39}                                        & 0.27                                                \\ 
\multicolumn{9}{l}{\textit{Open-weight Oracle Summary}}                                                                                                                                                                                                                                                                                                                                                                                                                                                                                \\
Qwen3.5-plus  & Qwen Code    & 54.16                                                                & 16.09                                                                    & 10.72                                                                & 21.54                                                    & 44.03                                                    & 1,668,167.12                                         & 455.18                                                 & \textbf{0.18}                                       \\
GLM-5    & Qwen Code         & 52.69                                                                & 15.32                                                                    & 10.75                                                                & 19.41                                                    & 41.39                                                    & 1,671,707.68                                         & 629.34                                                 & 0.48                                                \\
MiniMax 2.7   & Claude Code    & 49.87                                                                & 14.75                                                                    & 10.46                                                                & 22.07                                                    & 45.69                                                    & 1,229,219.56                                         & 420.92                                                 & 0.46                                               \\ \hline \hline

\end{tabular}
}\label{tab:all_base_agents}
\vspace{-11pt}
\end{table}

\subsubsection{Efficiency Analysis}
Efficiency trends show similar gains from summarized context, though the best model depends on the metric (Table~\ref{tab:all_base_agents}). GPT-5.3 Codex uses the fewest tool calls and tokens in both settings, making it the most efficient high-accuracy agent. Claude Sonnet~4.5 achieves the fastest runtime under oracle summaries at 280.62 seconds, suggesting summaries can significantly reduce search and editing time despite higher token usage. Among open-weight models, Qwen3.5-plus has the lowest monetary cost (\$0.18 with oracle summaries) but it still uses more tool calls and tokens than GPT-5.3 Codex. GLM-5 is consistently the slowest, exceeding 800s in the baseline, while Qwen3.5-plus has the highest token usage without matching gains in resolution. Overall, concise oracle guidance improves both effectiveness and efficiency, but trade-offs across runtime, tokens, and cost depend on the agent.

\begin{figure*}[hbt!]
\vspace{-5pt}
    \centering
    \includegraphics[width=0.97\linewidth]{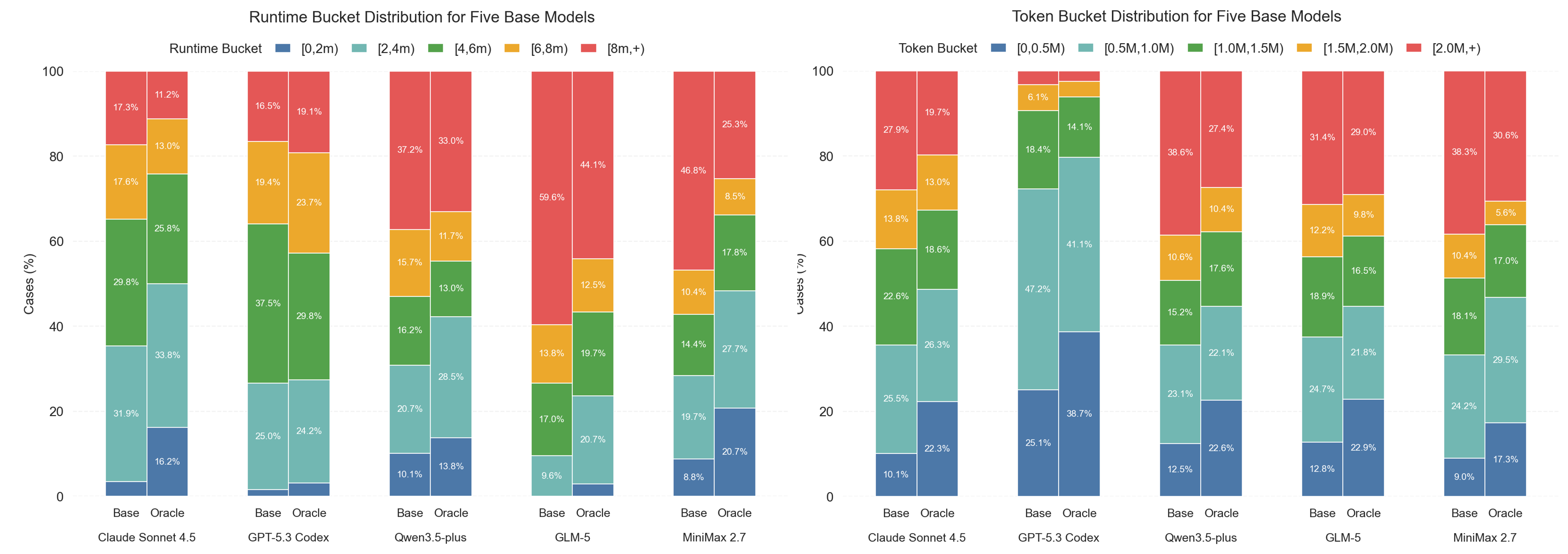} 
    \vspace{-5pt}
    \caption{Stacked distributions of runtime and token usage for five coding agents under the baseline and oracle summary settings. Each bar is binned into five ranges, and segment sizes indicate the percentage of tasks in each range.}
    \label{fig:efficiency}
\vspace{-10pt}
\end{figure*}

Figure~\ref{fig:efficiency} further illustrates these efficiency differences through the full distributions of runtime and token usage. Compared to baseline, oracle summaries generally shift runs toward lower runtime and token buckets, indicating that compact relevant context reduces both search overhead and repeated reasoning. This effect is strongest for Claude Sonnet~4.5 and GPT-5.3 Codex, whose distributions become more concentrated in the more efficient ranges, whereas GLM-5 and Qwen3.5-plus show greater variability. These results complement Table~\ref{tab:all_base_agents}, showing that summary reuse not only lowers average cost but also reduces especially expensive runs.

\subsection{Performance across Different Context Learning Settings}
We evaluate performance under five settings that differ in how prior context is exposed to the agent on the SWE-Bench Lite dataset with  Claude Sonnet 4.5. These settings allow a controlled comparison of prediction performance across different past context access and retrieval configurations.

\textbf{No-Context (Baseline)}: The agent is provided only with the repository name, base commit, and problem statement. It has no access to the context pool.

\textbf{Free Context Learning}: The agent has full access to the original context pool and can decide whether to retrieve and use any past context when solving the task.

\textbf{Oracle Context Learning}: The agent is explicitly provided with the most relevant past context trajectory, identified using the factual task relationships defined in Section~\ref{sec:task_instance_type}.

\textbf{Free Summary Learning}: We construct a \emph{summary context pool} by extracting the final summary section from each context trajectories. The agent is given full access to this \emph{summary context pool} and may decide whether and which summaries to use.

\textbf{Oracle Summary Learning}: The agent is explicitly given the summary of the most relevant past context, identified based on the known task relationships.

The summary files contain an average of 217.1 tokens, while the full context trajectories average 25,633.7 tokens, resulting in a substantially more compact representation of prior context.

\subsubsection{Accuracy Analysis}
Table~\ref{tab:pred_accuracy} reports six metrics. \texttt{Patch N/A} counts instances whose generated patches cannot be applied and are excluded from the remaining evaluation. \texttt{FAIL\_TO\_PASS} and \texttt{PASS\_TO\_PASS} are reported at both test and task levels, while \texttt{Resolved} is the overall SWE-ContextBench Lite resolution rate.

Oracle Summary Learning achieves the best accuracy, improving resolution from 26.26\% for the No-Context baseline to 34.34\%, while Oracle Context Learning provides a small gain to 27.27\%. The oracle--free gap is much larger for summaries than for full trajectories: Oracle Summary Learning outperforms Free Summary Learning by 12.12 points, while Oracle Context Learning exceeds Free Context Learning by only 1.01 points. Patch application failures are similar across settings, and all maintain high \texttt{PASS\_TO\_PASS} rates (92\%--95\%), indicating few regressions. Overall speaking, concise summaries are highly effective when correctly selected, but misleading when irrelevant.

\subsubsection{Efficiency Analysis}
For runtime, we evaluate all five settings on the 99 related tasks, yielding 495 runs in total. Oracle Context Learning is fastest on average at 5.95 minutes per task, followed by Free Summary Learning (6.28 min), the No-Context baseline (6.37 min), Oracle Summary Learning (6.66 min), and Free Context Learning (6.78 min). Median runtimes are tightly clustered between 331 and 356 seconds, implying that average differences are driven mainly by tail behavior. Free Context Learning shows the largest variance and worst-case runtime, exceeding 2,100 seconds, whereas Oracle Summary Learning yields the most stable runtime distribution. The benefit of oracle summaries is especially strong on harder tasks, where runtime falls by more than 60\% for the slowest instances.

For cost, token usage is dominated by cache-read tokens, which account for more than 97\% of total consumption across all settings. Oracle Context Learning has the lowest average cost at \$0.77 per instance, while Free Context Learning is the most expensive at \$0.98, a 27.3\% increase. The No-Context baseline remains close to Oracle Context Learning at \$0.79, whereas Free Summary Learning and Oracle Summary Learning cost \$0.91 and \$0.85, respectively. Thus, in this setup, summary-based reuse does not reduce cost relative to the baseline, and autonomous retrieval tends to increase repository access and overall token consumption.

\vspace{-5pt}
\begin{table}[t]
\centering
\caption{Comparison of different context learning settings and memory retrieval frameworks on 99 related tasks from SWE-ContextBench Lite. We report patch applicability, test-level and task-level pass rates for bug-fixing (\texttt{FAIL\_TO\_PASS}) and regression (\texttt{PASS\_TO\_PASS}) tests, overall task resolution rate, and efficiency in runtime and cost.}
\vspace{-4pt}
\resizebox{1\linewidth}{!}{
\begin{tabular}{c|cccccc|cc}
  \hline \hline
                                                                    & \multicolumn{6}{c|}{\textbf{Accuracy Analysis}}                                                                                                                                                                                                                                                                                                                                                                              & \multicolumn{2}{c}{\textbf{Efficiency Analysis}}                                                                             \\
Model                                                               & \begin{tabular}[c]{@{}c@{}}Patch N/A\\ (\%)\end{tabular} & \begin{tabular}[c]{@{}c@{}}FAIL\_TO\_PASS\\ Tests (\%)\end{tabular} & \begin{tabular}[c]{@{}c@{}}PASS\_TO\_PASS\\ Tests (\%)\end{tabular} & \begin{tabular}[c]{@{}c@{}}FAIL\_TO\_PASS\\ Tasks (\%)\end{tabular} & \begin{tabular}[c]{@{}c@{}}PASS\_TO\_PASS\\ Tasks (\%)\end{tabular} & \begin{tabular}[c]{@{}c@{}}Resolved\\ (\%)\end{tabular} & \begin{tabular}[c]{@{}c@{}}Time\\ (min)\end{tabular} & \begin{tabular}[c]{@{}c@{}}Cost\\ (\$)\end{tabular} \\ \hline
\begin{tabular}[c]{@{}c@{}}No-Context (Baseline)\end{tabular} & 3.03                                                     & 19.64                                                               & 94.91                                                               & 29.29                                                               & 88.89                                                               & 26.26                                                   & 6.37                                                 & 0.79                                                \\
\begin{tabular}[c]{@{}c@{}}Free Context Learning\end{tabular}     & 2.02                                                     & 22.22                                                               & 95.46                                                               & 28.28                                                               & 85.86                                                               & 26.26                                                   & 6.78                                                 & 0.98                                                \\
\begin{tabular}[c]{@{}c@{}}Oracle Context Learning\end{tabular}   & 4.04                                                     & 20.44                                                               & 94.41                                                               & 31.31                                                               & 89.90                                                               & 27.27                                                   & 5.95                                                 & 0.77                                                \\
\begin{tabular}[c]{@{}c@{}}Free Summary Learning\end{tabular}        & 7.07                                                     & 19.64                                                               & 93.41                                                               & 29.29                                                               & 86.87                                                               & 22.22                                                   & 6.28                                                 & 0.91                                                \\
\begin{tabular}[c]{@{}c@{}}Oracle Summary Learning\end{tabular}      & 6.06                                                     & 26.39                                                               & 92.53                                                               & 40.40                                                               & 85.86                                                               & 34.34                                                   & 6.66                                                 & 0.85                                                \\ \hline
Mem0     
& 14.14                                                    & 20.24                                                               & 82.44                                                               & 39.39                                                               & 86.87                                                               & 24.24                                                   & 4.72                                                 & 0.62   \\ 
OpenViking  
& 10.10                                                    & 57.34                                                               & 91.83                                                               & 40.40                                                               & 88.89                                                               & 29.20                                                   & 4.20                                                 & 0.53                                                \\
Supermemory    
& 10.10                                                    & 55.95                                                               & 90.23                                                               & 40.40                                                               & 88.89                                                               & 30.30                                                   & 5.04                                                 & 0.58                                                \\ \hline \hline  
\end{tabular}
}\label{tab:pred_accuracy}
\vspace{-13pt}
\end{table}


\subsection{Performance across Trending Memory Retrieval Frameworks} 
Table~\ref{tab:pred_accuracy} also compares four widely-used memory retrieval frameworks on 99 related tasks in SWE-ContextBench Lite: Mem0 \cite{chhikara2025mem0}, OpenViking \cite{openviking}, LangMem \cite{langmem2026}, and Supermemory \cite{supermemory}. Implementation details are provided in \ref{sec:appendix_memory_approaches_details}.

Supermemory performs best overall, achieving \texttt{FAIL\_TO\_PASS} test rate of 55.95\% and the highest resolution rate of 30.30\%, indicating superior capability in fixing previously failing tests. It also maintains strong \texttt{PASS\_TO\_PASS} scores (90.23\% at test level, 88.89\% at task level), suggesting that these gains do not come at the cost of regressions. OpenViking shows similar task-level performance (29.20\% resolved) and slightly higher \texttt{FAIL\_TO\_PASS} test accuracy (57.34\%), but without better overall outcomes. Mem0 underperforms across most metrics, with lower \texttt{FAIL\_TO\_PASS} scores (20.24\% at the test level, 39.39\% at the task level) and resolution rate (24.24\%), suggesting weaker context learning. In terms of efficiency, OpenViking demonstrates the fastest runtime (4.20 min) and lowest cost (\$0.53), while Supermemory remains competitive (5.04 min, \$0.58) given its stronger performance. Mem0 has higher cost (\$0.62) without corresponding accuracy gains. Overall, the results suggest that stronger memory use improves bug fixing effectiveness on related tasks, with Supermemory offering the best balance between performance and efficiency.

\begin{table}[t] 
\centering
\vspace{-5pt}
\caption{Retrieval quality of two context learning settings and four memory retrieval approaches on SWE-ContextBench Lite related tasks. For the top-1, top-2, and top-3 retrieved contexts, we report retrieval score, average context length, and match rate, together with the overall matching percentage.}
\vspace{-3pt}
\resizebox{0.95\linewidth}{!}{
\begin{tabular}{c|c|ccc|ccc|ccc|c}
            \hline \hline
            & \multirow{2}{*}{\textbf{\begin{tabular}[c]{@{}c@{}}Number of \\ Retrieved\\ Contexts\end{tabular}}} & \multicolumn{3}{c|}{\textbf{Top1 Retrieved Context}}                                                                                                                 & \multicolumn{3}{c|}{\textbf{Top2 Retrieved Context}}                                                                                                                 & \multicolumn{3}{c|}{\textbf{Top3 Retrieved Context}}                                                                                                                 & \textbf{Overall}                                       \\
            &                                                                                                     & \begin{tabular}[c]{@{}c@{}}Score\\ (\%)\end{tabular} & \begin{tabular}[c]{@{}c@{}}Avg\\ Tokens\end{tabular} & \begin{tabular}[c]{@{}c@{}}Matched\\ (\%)\end{tabular} & \begin{tabular}[c]{@{}c@{}}Score\\ (\%)\end{tabular} & \begin{tabular}[c]{@{}c@{}}Avg\\ Tokens\end{tabular} & \begin{tabular}[c]{@{}c@{}}Matched\\ (\%)\end{tabular} & \begin{tabular}[c]{@{}c@{}}Score\\ (\%)\end{tabular} & \begin{tabular}[c]{@{}c@{}}Avg\\ Tokens\end{tabular} & \begin{tabular}[c]{@{}c@{}}Matched\\ (\%)\end{tabular} & \begin{tabular}[c]{@{}c@{}}Matched\\ (\%)\end{tabular} \\ \hline
Free Context Learning       &      Self-Determine                                              &       -                                         &         136,907                                       &         18.18                                         &            -                                     &            78,044                                     &                  0                                 &    -                                             &         77,129                                        &           0                                         &      18.18                                             \\
Free Summary Learning       &      Self-Determine                                                                                              &        -                                        &            372.79                                    &          34.34                                        &   -                                              &    363.78                                             &      2.02                                             &        -                                         &         368.33                                        &         0                                           &      36.36                                             \\ 
Mem0 
& 3                                                                                                   & 55.91                                                & 11.97                                                & 25.25                                                  & 50.54                                                & 11.15                                                & 15.15                                                  & 47.55                                                & 11.55                                                & 9.09                                                   & 39.39                                                  \\
OpenViking 
& 3                                                                                                   & 50.40                                                & 245.90                                               & 38.38                                                  & 48.95                                                & 254.16                                               & 35.35                                                  & 48.13                                                & 249.33                                               & 40.40                                                  & 51.52                                                  \\
LangMem 
& 10                                                                                                  & 49.70                                                & 48.18                                                & 55.56                                                  & 45.51                                                & 47.84                                                & 35.35                                                  & 42.65                                                & 48.42                                                & 14.14                                                  & 73.34                 \\ 
Supermemory 
& 15                                                                                                  & 74.51                                                & 30.06                                                & 23.23                                                  & 72.76                                                & 28.86                                                & 20.20                                                  & 71.84                                                & 26.62                                                & 11.11                                                  & 59.60                 \\ \hline \hline                                       
\end{tabular}
}\label{tab:matched_analysis}
\vspace{-10pt}
\end{table}
Table~\ref{tab:matched_analysis} compares two context learning settings and four memory approaches in how they transform trajectories into retrievable contexts and how often the retrieved contexts match the gold related experience task. Among context learning settings, Free Context Learning retrieves extremely long contexts (136k tokens for top-1) but achieves low matching (18.18\%). Free Summary Learning uses much shorter contexts (\textasciitilde 370 tokens) and improves matching to 36.36\%, showing the benefit of compact summaries. Among memory retrieval frameworks, LangMem achieves the highest overall matching rate of 73.34\%, with strong top-1 matching of 55.56\%. Supermemory maintains a high overall match rate at 59.60\% with compact contexts (\textasciitilde 30 tokens). OpenViking retrieves much longer contexts, averaging 245.90--254.16 tokens, and achieves overall matching rate at 51.52\%. In contrast, Mem0 is the most compact method but performs worst, with a 39.39\% match rate. These results suggest that more compact and structured context representations improve both efficiency and retrieval quality, with methods like LangMem and Supermemory achieving the strongest matching performance.

\section{Related Work}
Recent work on LLM memory focuses on long-context understanding and persistent memory in general domains. Benchmarks such as LongBench~\cite{bai2024longbench}, LongMemEval~\cite{wu2024longmemeval}, RULER~\cite{hsieh2024ruler}, LoCoMo~\cite{maharana2024evaluating}, and PerLTQA~\cite{du2024perltqa} evaluate abilities such as information extraction, multi-session reasoning, and knowledge update over documents, conversations, or synthetic contexts. More agent-oriented benchmarks, such as MemoryRewardBench~\cite{tang2026texttt} and MemBench~\cite{tan2025membench}, move toward interactive memory use but remain focused on general domains rather than repository-level software engineering.

In parallel, coding benchmarks like RepoBench~\cite{liu2306repobench}, SWE-Bench~\cite{jimenez2024swebench}, LongCodeBench~\cite{rando2025longcodebench}, and LoCoBench~\cite{qiu2025locobench} evaluate retrieval, code understanding, and bug fixing in realistic repositories. While they provide repository-level context, they do not evaluate whether an agent can learn from prior problem-solving contexts across tasks. Our benchmark differs in that it treats past trajectories as retrievable context, including reasoning steps, tool use, and intermediate decisions, thereby enabling direct evaluation of cross-task context learning. Table~\ref{tab:memory_benchmarks} summarizes these differences.

\begin{table}[H]
\centering
\vspace{-4pt}
\caption{Comparison of general memory and coding benchmarks across four abilities: information extraction (IE), multi-session reasoning (MR), knowledge update (KU), and cross-task context learning (CT). Existing coding benchmarks focus on repository-level context, while SWE-ContextBench introduces trajectory-level memory and directly evaluates cross-task context learning.}
\vspace{-4pt}
\resizebox{0.98\linewidth}{!}{
\small
\begin{tabular}{lccccccccc}
\hline \hline
\multirow{2}{*}{\textbf{Method}} &
\multirow{2}{*}{\textbf{Domain}} &
\multirow{2}{*}{\textbf{Memory}} &
\multirow{2}{*}{\textbf{Access}} &
\multirow{2}{*}{\textbf{Agent}} &
\multirow{2}{*}{\textbf{Key Property}} &
\multicolumn{4}{c}{\textbf{Core Memory Abilities}} \\
\cmidrule(lr){7-10}
 &  &  &  &  &  & \textbf{IE} & \textbf{MR} & \textbf{KU} & \textbf{CT} \\
\midrule

\multicolumn{10}{l}{\textit{General memory benchmarks}} \\
LongBench~\cite{bai2024longbench} & General & Document & Full & $\times$ & Context processing & $\checkmark$ & $\checkmark$ & $\times$ & $\times$ \\
LongMemEval~\cite{wu2024longmemeval} & General & Document & Full & $\times$ & Long-term reasoning & $\checkmark$ & $\checkmark$ & $\checkmark$ & $\times$ \\
RULER~\cite{hsieh2024ruler} & General & Synthetic context & Full & $\times$ & Retrieval probing & $\checkmark$ & $\times$ & $\times$ & $\times$ \\
LoCoMo~\cite{maharana2024evaluating} & Dialogue & Conversation & Full & $\times$ & Episodic memory & $\checkmark$ & $\checkmark$ & $\times$ & $\times$ \\
PerLTQA~\cite{du2024perltqa} & QA & Document & Full & $\times$ & Long-term QA & $\checkmark$ & $\times$ & $\times$ & $\times$ \\
MemoryRewardBench~\cite{tang2026texttt} & General & Mixed & Policy & $\times$ & Policy learning & $\times$ & $\times$ & $\checkmark$ & $\times$ \\
BEAM~\cite{tavakoli2025beyond} & General & Document & Full & $\times$ & Ultra-long context & $\checkmark$ & $\checkmark$ & $\times$ & $\times$ \\
MemBench~\cite{tan2025membench} & General & Interaction & Retrieval & $\checkmark$ & Memory interaction & $\checkmark$ & $\checkmark$ & $\checkmark$ & $\times$ \\

\midrule
\multicolumn{10}{l}{\textit{Coding benchmarks}} \\
RepoBench~\cite{liu2306repobench} & Coding & Repository & Full & $\times$ & Retrieval, Completion and Pipeline & $\checkmark$ & $\checkmark$ & $\times$ & $\times$ \\
SWE-Bench~\cite{jimenez2024swebench} & Coding & Repository & Full & $\checkmark$ & Repository-level bug fixing & $\checkmark$ & $\checkmark$ & $\times$ & $\times$ \\
LongCodeBench~\cite{rando2025longcodebench} & Coding & Repository & Full & $\checkmark$ & Code Comprehension and Repair & $\checkmark$ & $\checkmark$ & $\times$ & $\times$ \\
LoCoBench~\cite{qiu2025locobench} & Coding & Repository & Full & $\times$ & Complex long-context software engineering & $\checkmark$ & $\checkmark$ & $\times$ & $\times$ \\
\textbf{SWE-ContextBench} & Coding & Trajectory & Retrieval & $\checkmark$ & \textbf{Cross-task context learning} & $\checkmark$ & $\checkmark$ & $\checkmark$ & $\checkmark$ \\ \hline \hline

\end{tabular}
}
\label{tab:memory_benchmarks}
\vspace{-10pt}
\end{table}

\section{Conclusion}
We introduces SWE-ContextBench, a benchmark for studying context learning in repository-level coding agents. Built from SWE-Bench Lite, SWE-Bench Multilingual, and SWE-Bench Verified, SWE-ContextBench contains 1,476 tasks from 51 repositories and 9 programming languages, including 1,100 base tasks and 376 related tasks capturing explicit cross-task dependencies. Our experiments show that context learning is beneficial only when the reused context is both compact and correctly selected. Oracle Summary Learning achieves the highest task resolution, while Oracle Context Learning yields the lowest average runtime and cost. In contrast, free retrieval often brings limited gains and can even hurt performance, especially when context are irrelevant. Results across five coding agents further show that both model capability and reuse strategy jointly shape the trade-off between effectiveness and efficiency, while experiments with memory retrieval frameworks confirm that stronger retrieval quality improves downstream performance on related tasks. More insights to improve context learning in coding agents are discussed in \ref{sec:appendix_insights.}. Overall, SWE-ContextBench complements existing coding and memory benchmarks by providing a realistic and controlled testbed for studying how agents retrieve, represent, and reuse prior contexts in software engineering tasks.


\bibliographystyle{unsrtnat}
\bibliography{custom}

@inproceedings{
    jimenez2024swebench,
    title={{SWE}-bench: Can Language Models Resolve Real-world Github Issues?},
    author={Carlos E Jimenez and John Yang and Alexander Wettig and Shunyu Yao and Kexin Pei and Ofir Press and Karthik R Narasimhan},
    booktitle={The Twelfth International Conference on Learning Representations},
    year={2024},
    url={https://openreview.net/forum?id=VTF8yNQM66}
    }

@article{wong2025widesearch,
  title={Widesearch: Benchmarking agentic broad info-seeking},
  author={Wong, Ryan and Wang, Jiawei and Zhao, Junjie and Chen, Li and Gao, Yan and Zhang, Long and Zhou, Xuan and Wang, Zuo and Xiang, Kai and Zhang, Ge and others},
  journal={arXiv preprint arXiv:2508.07999},
  year={2025}
}

@inproceedings{mialon2023gaia,
  title={Gaia: a benchmark for general ai assistants},
  author={Mialon, Gr{\'e}goire and Fourrier, Cl{\'e}mentine and Wolf, Thomas and LeCun, Yann and Scialom, Thomas},
  booktitle={The Twelfth International Conference on Learning Representations},
  year={2023}
}

@article{wei2025browsecomp,
  title={Browsecomp: A simple yet challenging benchmark for browsing agents},
  author={Wei, Jason and Sun, Zhiqing and Papay, Spencer and McKinney, Scott and Han, Jeffrey and Fulford, Isa and Chung, Hyung Won and Passos, Alex Tachard and Fedus, William and Glaese, Amelia},
  journal={arXiv preprint arXiv:2504.12516},
  year={2025}
}

@article{xu2025mem,
  title={A-mem: Agentic memory for llm agents},
  author={Xu, Wujiang and Liang, Zujie and Mei, Kai and Gao, Hang and Tan, Juntao and Zhang, Yongfeng},
  journal={arXiv preprint arXiv:2502.12110},
  year={2025}
}

@article{chhikara2025mem0,
  title={Mem0: Building production-ready ai agents with scalable long-term memory},
  author={Chhikara, Prateek and Khant, Dev and Aryan, Saket and Singh, Taranjeet and Yadav, Deshraj},
  journal={arXiv preprint arXiv:2504.19413},
  year={2025}
}

@misc{openviking,
  author       = {{Volcengine Viking Team}},
  title        = {OpenViking: An Open-Source Context Database for AI Agents},
  year         = {2026},
  publisher    = {GitHub},
  journal      = {GitHub repository},
  howpublished = {\url{https://github.com/volcengine/OpenViking}},
  note         = {Provides a filesystem-based paradigm for managing agent memory, resources, and skills},
}

@misc{langmem2026,
  author       = {{LangChain AI}},
  title        = {LangMem: Long-Term Memory for LLM Agents},
  year         = {2026},
  publisher    = {GitHub},
  journal      = {GitHub repository},
  howpublished = {\url{https://github.com/langchain-ai/langmem}},
}

@misc{supermemory,
  author       = {{Supermemory AI}},
  title        = {Supermemory: Memory and Context Layer for AI},
  year         = {2026},
  publisher    = {GitHub},
  journal      = {GitHub repository},
  howpublished = {\url{https://github.com/supermemoryai/supermemory}},
}

@article{li2025memos,
  title={MemOS: An Operating System for Memory-Augmented Generation (MAG) in Large Language Models},
  author={Li, Zhiyu and Song, Shichao and Wang, Hanyu and Niu, Simin and Chen, Ding and Yang, Jiawei and Xi, Chenyang and Lai, Huayi and Zhao, Jihao and Wang, Yezhaohui and others},
  journal={arXiv preprint arXiv:2505.22101},
  year={2025}
}

@article{yang2024swe,
  title={Swe-agent: Agent-computer interfaces enable automated software engineering},
  author={Yang, John and Jimenez, Carlos E and Wettig, Alexander and Lieret, Kilian and Yao, Shunyu and Narasimhan, Karthik and Press, Ofir},
  journal={Advances in Neural Information Processing Systems},
  volume={37},
  pages={50528--50652},
  year={2024}
}

@article{zeng2025glm,
  title={Glm-4.5: Agentic, reasoning, and coding (arc) foundation models},
  author={Zeng, Aohan and Lv, Xin and Zheng, Qinkai and Hou, Zhenyu and Chen, Bin and Xie, Chengxing and Wang, Cunxiang and Yin, Da and Zeng, Hao and Zhang, Jiajie and others},
  journal={arXiv preprint arXiv:2508.06471},
  year={2025}
}

@article{zhou2025mem1,
  title={MEM1: Learning to Synergize Memory and Reasoning for Efficient Long-Horizon Agents},
  author={Zhou, Zijian and Qu, Ao and Wu, Zhaoxuan and Kim, Sunghwan and Prakash, Alok and Rus, Daniela and Zhao, Jinhua and Low, Bryan Kian Hsiang and Liang, Paul Pu},
  journal={arXiv preprint arXiv:2506.15841},
  year={2025}
}

@article{yu2025memagent,
  title={Memagent: Reshaping long-context llm with multi-conv rl-based memory agent},
  author={Yu, Hongli and Chen, Tinghong and Feng, Jiangtao and Chen, Jiangjie and Dai, Weinan and Yu, Qiying and Zhang, Ya-Qin and Ma, Wei-Ying and Liu, Jingjing and Wang, Mingxuan and others},
  journal={arXiv preprint arXiv:2507.02259},
  year={2025}
}

@inproceedings{chhikara2023knowledge,
  title={Knowledge-enhanced agents for interactive text games},
  author={Chhikara, Prateek and Zhang, Jiarui and Ilievski, Filip and Francis, Jonathan and Ma, Kaixin},
  booktitle={Proceedings of the 12th Knowledge Capture Conference 2023},
  pages={157--165},
  year={2023}
}

@article{liu2306repobench,
  title={Repobench: Benchmarking repository-level code auto-completion systems, 2023},
  author={Liu, Tianyang and Xu, Canwen and McAuley, Julian},
  journal={URL https://arxiv. org/abs/2306.03091}
}

@inproceedings{jimenez2023swe,
  title={Swe-bench: Can language models resolve real-world github issues?},
  author={Jimenez, Carlos E and Yang, John and Wettig, Alexander and Yao, Shunyu and Pei, Kexin and Press, Ofir and Narasimhan, Karthik R},
  booktitle={The twelfth international conference on learning representations},
  year={2023}
}

@article{chen2021evaluating,
  title={Evaluating large language models trained on code},
  author={Chen, Mark and Tworek, Jerry and Jun, Heewoo and Yuan, Qiming and Pinto, Henrique Ponde De Oliveira and Kaplan, Jared and Edwards, Harri and Burda, Yuri and Joseph, Nicholas and Brockman, Greg and others},
  journal={arXiv preprint arXiv:2107.03374},
  year={2021}
}

@article{austin2021program,
  title={Program synthesis with large language models},
  author={Austin, Jacob and Odena, Augustus and Nye, Maxwell and Bosma, Maarten and Michalewski, Henryk and Dohan, David and Jiang, Ellen and Cai, Carrie and Terry, Michael and Le, Quoc and others},
  journal={arXiv preprint arXiv:2108.07732},
  year={2021}
}

@article{rando2025longcodebench,
  title={Longcodebench: Evaluating coding llms at 1m context windows},
  author={Rando, Stefano and Romani, Luca and Sampieri, Alessio and Franco, Luca and Yang, John and Kyuragi, Yuta and Galasso, Fabio and Hashimoto, Tatsunori},
  journal={arXiv preprint arXiv:2505.07897},
  year={2025}
}

@article{qiu2025locobench,
  title={Locobench: A benchmark for long-context large language models in complex software engineering},
  author={Qiu, Jielin and Liu, Zuxin and Liu, Zhiwei and Murthy, Rithesh and Zhang, Jianguo and Chen, Haolin and Wang, Shiyu and Zhu, Ming and Yang, Liangwei and Tan, Juntao and others},
  journal={arXiv preprint arXiv:2509.09614},
  year={2025}
}

@inproceedings{bai2024longbench,
  title={Longbench: A bilingual, multitask benchmark for long context understanding},
  author={Bai, Yushi and Lv, Xin and Zhang, Jiajie and Lyu, Hongchang and Tang, Jiankai and Huang, Zhidian and Du, Zhengxiao and Liu, Xiao and Zeng, Aohan and Hou, Lei and others},
  booktitle={Proceedings of the 62nd annual meeting of the association for computational linguistics (volume 1: Long papers)},
  pages={3119--3137},
  year={2024}
}

@article{wu2024longmemeval,
  title={Longmemeval: Benchmarking chat assistants on long-term interactive memory},
  author={Wu, Di and Wang, Hongwei and Yu, Wenhao and Zhang, Yuwei and Chang, Kai-Wei and Yu, Dong},
  journal={arXiv preprint arXiv:2410.10813},
  year={2024}
}

@article{hsieh2024ruler,
  title={RULER: What's the real context size of your long-context language models?},
  author={Hsieh, Cheng-Ping and Sun, Simeng and Kriman, Samuel and Acharya, Shantanu and Rekesh, Dima and Jia, Fei and Zhang, Yang and Ginsburg, Boris},
  journal={arXiv preprint arXiv:2404.06654},
  year={2024}
}

@inproceedings{maharana2024evaluating,
  title={Evaluating very long-term conversational memory of llm agents},
  author={Maharana, Adyasha and Lee, Dong-Ho and Tulyakov, Sergey and Bansal, Mohit and Barbieri, Francesco and Fang, Yuwei},
  booktitle={Proceedings of the 62nd Annual Meeting of the Association for Computational Linguistics (Volume 1: Long Papers)},
  pages={13851--13870},
  year={2024}
}

@inproceedings{du2024perltqa,
  title={Perltqa: A personal long-term memory dataset for memory classification, retrieval, and fusion in question answering},
  author={Du, Yiming and Wang, Hongru and Zhao, Zhengyi and Liang, Bin and Wang, Baojun and Zhong, Wanjun and Wang, Zezhong and Wong, Kam-Fai},
  booktitle={Proceedings of the 10th SIGHAN Workshop on Chinese Language Processing (SIGHAN-10)},
  pages={152--164},
  year={2024}
}

@article{tang2026texttt,
  title={{\texttt{MemoryRewardBench}: Benchmarking Reward Models for Long-Term Memory Management in Large Language Models}},
  author={Tang, Zecheng and Ji, Baibei and Sun, Ruoxi and Wang, Haitian and You, WangJie and Yijun, Zhang and Zhu, Wenpeng and Qi, Ji and Li, Juntao and Zhang, Min},
  journal={arXiv preprint arXiv:2601.11969},
  year={2026}
}

@article{tavakoli2025beyond,
  title={Beyond a million tokens: Benchmarking and enhancing long-term memory in llms},
  author={Tavakoli, Mohammad and Salemi, Alireza and Ye, Carrie and Abdalla, Mohamed and Zamani, Hamed and Mitchell, J Ross},
  journal={arXiv preprint arXiv:2510.27246},
  year={2025}
}

@inproceedings{tan2025membench,
  title={Membench: Towards more comprehensive evaluation on the memory of llm-based agents},
  author={Tan, Haoran and Zhang, Zeyu and Ma, Chen and Chen, Xu and Dai, Quanyu and Dong, Zhenhua},
  booktitle={Findings of the Association for Computational Linguistics: ACL 2025},
  pages={19336--19352},
  year={2025}
}

\newpage
\appendix

\section{Technical appendices and supplementary material}
\subsection{Prompt Template Example}\label{sec:appendix_instruction}
We provide the coding agents following instructions to run each task:
\begin{Verbatim}[fontsize=\small]
instance:
  instance_id: <instance_id>
  repo: <repo>
  base_commit: <base_commit>
  problem_statement: <problem_statement>

setup:
  testbed_dir: testbed

  steps:
    - description: Remove existing testbed if present
      command: rm -rf testbed

    - description: Clone repository
      command: git clone https://github.com/<repo>.git testbed

    - description: Checkout base commit
      command: git -C testbed fetch origin <base_commit> && \
      git -C testbed checkout <base_commit>

    - description: Install dependencies
      command: pip install -e testbed

    - description: Remove git history to prevent solution leakage
      command: rm -rf testbed/.git

    - description: Re-initialize git without history
      command: git init testbed

output:
  directory: output
  filename: <instance_id>_preds.json

  format:
    model_name_or_path: <auto_detect>
    instance_id: <instance_id>
    model_patch: <generated_patch>

  patch_generation:
    - command: git -C testbed add -A
    - command: git -C testbed diff --cached

cleanup:
  - description: Remove testbed after saving preds
    command: rm -rf testbed

instructions: |
  1. Run setup steps to prepare the testbed
  2. Navigate to testbed/ and fix the bug described in problem_statement
  3. You should not include tests in the generated patch.
  4. Generate patch: git -C testbed add -A && git -C testbed diff --cached
  5. Save to output/<instance_id>_preds.json with format:
     {
       "<instance_id>": {
         "model_name_or_path": "<auto_detect>",
         "instance_id": "<instance_id>",
         "model_patch": "<patch>"
       }
     }
  6. Remove testbed folder: rm -rf testbed
\end{Verbatim}

\subsection{Implementation Details for Memory Retrieval Frameworks}\label{sec:appendix_memory_approaches_details}
In this section, we describe the implementation details of four memory retrieval frameworks: Mem0 \cite{chhikara2025mem0}, OpenViking \cite{openviking}, LangMem \cite{langmem2026}, and Supermemory \cite{supermemory}. To incorporate prior context, we use the collected context trajectory files as input, as described in Section~\ref{sec:experience_traj}.

\subsubsection{Mem0}
\textbf{Memory Ingestion.} Previous context trajectories, stored as JSONL conversation files, are preprocessed and ingested into a persistent local memory store. Each file is parsed to extract a structured representation comprising a high-level summary, the complete user/assistant message history, and associated metadata (repository name, instance ID). Each context is then committed to a local Mem0 Memory instance via two sequential add calls: the first stores a concise summary concatenated with the problem statement, and the second stores the full conversation transcript. Both calls are tagged with a unique run ID and structured metadata to enable downstream filtering. Following the standard Mem0 interface, infer=True is set on both calls, instructing an LLM (GPT-4o-mini) to extract and deduplicate salient facts prior to storage. The resulting embeddings are persisted in a local Qdrant vector store, with memory metadata recorded in a co-located SQLite database. Ingesting 300 trajectories requires approximately 21 minutes at a total cost of \$0.24.  

\textbf{Memory Retrieval.} At inference time, relevant past contexts are retrieved via semantic similarity search over the ingested vector store. Given a natural-language query describing the current problem, an embedding is computed using OpenAI's text-embedding-3-small model and issued directly against the Qdrant collection. This enables cross-session retrieval across the full corpus of ingested contexts. Retrieved candidates are ranked by cosine similarity score and filtered by a type: past\_context predicate, with an optional repository-level filter to narrow scope. The top-$k$ results, comprising memory text and associated metadata, are serialized to a structured JSON output file for consumption by downstream components.

\subsubsection{OpenViking}
 \textbf{Memory Ingestion.} Past context trajectories are ingested into OpenViking by registering each JSONL file as a resource via \texttt{client.add\_resource()}, which triggers OpenViking's internal indexing pipeline to parse, chunk, and embed the file contents automatically. Files are processed in batches of 25, with each batch followed by a synchronous \texttt{client.wait\_processed()} call (up to 100 minutes per batch) that blocks until all embeddings and indexes for that batch are confirmed complete. This process requires no manual extraction of summaries or message structure — OpenViking handles document parsing and semantic indexing natively. Ingesting 300 trajectories required approximately 570 minutes at a total cost of \$18.96.                       
 
 \textbf{Memory Retrieval.} At inference time, relevant past contexts are retrieved by issuing a \texttt{client.find()} query against the full resource collection (viking://resources/), using OpenViking's built-in semantic search with a relevance score threshold of 0.2. For each result above the threshold, the resource URI, relevance score, and auto-generated abstract are returned; the full trajectory content is then fetched via a separate \texttt{client.read()} call. The top-$k$ results are serialized to a structured JSON output file for downstream consumption.

\subsubsection{LangMem}
\textbf{Memory Ingestion.} Each JSONL trajectory is parsed to reconstruct the full conversation as an ordered list of role/content messages (user, assistant, system), with session metadata (working directory, session ID) prepended as a system message and tool calls normalized to inline text. The complete message list is then passed to LangMem's \texttt{create\_memory\_manager} --- backed by GPT-4o-mini with a structured \texttt{Memory} schema --- which invokes the LLM to extract discrete, self-contained factual memory items from the conversation. Each extracted memory is stored individually in a persistent LangGraph \texttt{SqliteStore} with a vector index (\texttt{text-embedding-3-small}) under a shared \texttt{memories} namespace. Ingesting 300 trajectories yielded 997 memories in approximately 58 minutes at a cost of \$0.98.                                    
            
\textbf{Memory Retrieval.} At inference time, relevant memories are retrieved  by issuing a \texttt{store.search()} call against the SQLite vector index. The query string is embedded using \texttt{text-embedding-3-small} and matched against all stored memory embeddings in the \texttt{memories} namespace. The top-$k$ results (default 10) are returned with relevance scores and serialized to a structured JSON output file for downstream consumption. 

\subsubsection{Supermemory}
\textbf{Memory Ingestion.} Each JSONL trajectory is parsed and formatted into human-readable conversation text by extracting user and assistant turns, then uploaded to Supermemory's managed cloud platform via \texttt{client.add()}, tagged under the \texttt{past\_context} container with per-file metadata (filename, session ID, message count). Content extraction, embedding, and RAG indexing are handled entirely by Supermemory's closed-source backend. A 1.5-second rate-limit delay is applied between uploads. Ingesting 300 trajectories consumed 369.6k tokens and completed in under one hour.

\textbf{Memory Retrieval.} Supermemory exposes four MCP tools: \texttt{mcp\_\_supermemory\_\_memory} (save or forget a piece of information to/from the memory store), \texttt{mcp\_\_supermemory\_\_recall} (search saved memories using a query to retrieve relevant past context), \texttt{mcp\_\_supermemory\_\_listProjects} (list all available projects/containers used to organize memories), and \texttt{mcp\_\_supermemory\_\_whoAmI} (return current logged-in user info). At inference time, relevant past contexts are retrieved via \texttt{mcp\_\_supermemory\_\_recall}, which performs semantic search over the \texttt{past\_context} container. 

Specifically, we provided the following instruction to the agent: 
\begin{Verbatim}[fontsize=\small]
  When asked to fix a bug, always call mcp__supermemory__recall first with                                           
  the bug description and containerTag: "past_context" to search for  
  past similar bugs and relevant context before attempting a fix. If                                           
  relevant past context is found, learn from it to solve the new issue.   
  Do not save the new issue to memory.                                                  
\end{Verbatim}

\subsection{Insights to Improve Context Learning in Coding Agents} \label{sec:appendix_insights.}
\subsubsection{Summarization Quality}
Results in Table~\ref{tab:all_base_agents} show that effective context learning depends on concise and informative summaries of long trajectories. Raw context logs are often too verbose and noisy, making them hard to reuse directly. Instead, agents benefit from compact representations that capture key insights. This highlights the importance of strong summarization ability. For example, Mem0 retrieves a brief description (“Problem addressed involves limiting choices on a ForeignKey that can render duplicate options in formfield”), while LangMem produces a more precise and detailed summary (“The bug identified by django\_\_django-13315 involves limit\_choices\_to where using a Q object could lead to duplicate entries in Django formfield dropdowns due to SQL joins producing multiple rows for the same object if there are multiple matching related objects.”), which better captures the root cause. As a result, richer summaries provide more useful guidance and lead to stronger performance.

\subsubsection{Context Management}
Current approaches treat stored context as an unstructured pool, without organization or indexing. However, efficient systems (e.g., file systems) rely on structured storage for fast retrieval. Similarly, organizing contexts—for example, grouping by repository, task type, or dependency structure—could significantly improve retrieval efficiency and relevance. A hierarchical or repository-level organization may help agents locate useful context more effectively.

\subsubsection{Adaptive Learning Decisions}
Table~\ref{tab:pred_accuracy} shows that naive context learning can increase cost without improving accuracy: both Free Context Learning and Free Summary Learning are more expensive than the no-context baseline. This suggests that learning is not always beneficial. For simpler tasks, solving from scratch may be more efficient, while for complex tasks, prior context can reduce computation. Therefore, agents should learn to decide when to reuse context and when to ignore it, balancing cost and benefit.

\subsubsection{Retrieval Strategy}
Results in Table~\ref{tab:pred_accuracy} and Table~\ref{tab:matched_analysis} show that memory retrieval frameworks improve performance by retrieving relevant past cases. However, most approaches use a fixed number of retrieved contexts, which may not be suitable for all tasks. Some problems benefit from multiple related examples to capture different solution patterns, while others require only a single highly relevant case. This suggests that retrieval should be adaptive, with the agent selecting both the number and relevance of contexts based on task complexity. In addition, stronger retrieval strategies that improve matching quality are important. Higher matching scores indicate better identification of truly related cases, which allows agents to learn more effectively from prior context. Improving both the precision and flexibility of retrieval is therefore key to maximizing the benefits of context learning.

\subsubsection{Utilization of Retrieved Context}
Even when the correct prior context is identified, agents may still fail to solve the task. Past trajectories often include errors, trial-and-error steps, or incomplete reasoning that were only corrected later in the process. If an agent blindly follows these intermediate steps, it can be misled and reproduce incorrect decisions. This highlights that retrieval alone is not sufficient. Agents must actively interpret the retrieved context, distinguishing between useful insights and misleading details. In practice, this means identifying the core cause of the issue, extracting the relevant solution pattern, and adapting it to the current task rather than copying it directly. It also requires the ability to disregard irrelevant or incorrect parts of the trajectory. Therefore, effective context learning depends not only on retrieving the right context, but also on using it critically and selectively. Strengthening this ability to evaluate, adapt, and filter past context is essential for achieving robust and reliable performance.

\subsection{Context Pool Quality and Noise}\label{sec:appendex_experience_pool_quality}
The quality of the context pool is not explicitly controlled or filtered. We do not explicitly evaluate whether the \textit{experience tasks} are correctly solved or fully resolved. This design choice aims to reflect real-world software engineering environments, where solutions often involve multiple iterations, partial fixes, and unresolved issues. In practice, development processes include trial-and-error, internal corrections, and incomplete outcomes, all of which are naturally captured in the recorded trajectories. As a result, the context pool may contain trajectories that are partially correct, incorrect, or include residual bugs. We intentionally do not filter out such cases, as they represent realistic development conditions. However, this also means that agents must learn to handle noisy and imperfect context, which can make effective reuse more challenging.

\subsection{Limitations}
A key limitation of SWE-ContextBench is the lack of comprehensive human validation. While related task instances are manually identified and verified for structural correctness, there is no systematic involvement of human annotators to assess the clarity of problem descriptions, the correctness of test patches, or the overall solvability of each task. As a result, some instances may contain noise, especially for complex or indirectly linked cases derived from issue and pull request relationships.


\end{document}